\documentclass[%
aps,
 amsmath,amssymb,
 reprint,%
]{revtex4-1}
\usepackage{graphicx}
\usepackage{dcolumn}
\usepackage{bm}
\begin{document}

\preprint{AIP/123-QED}

\title{Splitting up entropy into vibrational and configurational contributions\\ in bulk metallic glasses:
	A thermodynamic approach}

\author{Ren\'e Alvarez-Donado}
\email{ralvarez@ifi.unicamp.br}
\author{Alex Antonelli}
 \email{aantone@ifi.unicamp.br}
  \affiliation{ 
  	Instituto de F\'isica Gleb Wataghin and Center for Computing in Engineering \& Sciences, Universidade Estadual de Campinas, UNICAMP, 13083-859 Campinas, S\~ao Paulo, Brazil 
  }%
\begin{abstract}
We applied an efficient methodology to separate vibrational and configurational entropies in bulk metallic glasses by means of molecular dynamics simulation based on a combination of non-equilibrium adiabatic switching and reversible scaling methods. This approach involves calculating the vibrational free energy using the Einstein crystal as a reference for the solid phase and the recently proposed Uhlenbeck-Ford model for the fluid phase. This methodology has the advantage that it does not require a crystalline solid phase for separating the entropies. Therefore, in principle, it is applicable to any material, regardless of whether or not it has a crystalline phase. Using this methodology, we separate the vibrational and configurational entropies of two metallic glasses with different fragilities at zero external pressure, namely, Cu$_{50}$Zr$_{50}$ and  Cu$_{46}$Zr$_{46}$Al$_{8}$. We find that the results for the former alloy are in quite reasonable agreement with recent experimental work by Smith \textit{et al.}[H. L. Smith \textit{et al.}, Nat. Phys. \textbf{13}, 900 (2017)]. We also find the configurational entropy of the glass containing Al to be 70\% larger than that of the other glass. Our results suggest that, although other factors may be at play, the configurational entropy can be used to investigate the effect of the addition of a minor-alloying element on the glass-forming ability of bulk metallic glasses.
\end{abstract}

\maketitle

\section{INTRODUCTION}
The excess entropy in glass-forming liquids in relation to the crystalline phase is a phenomenon studied since the early 1900's. It started with measurements of the caloric properties of glass-forming substances done by Nernst in order to verify whether or not the third law of thermodynamics was applicable to all forms of condensed matter\cite{Schiller2007}, which was followed by the publication in 1926 by Simon and Lange of their results about finite values of entropy for $T$ = 0 K in glasses of glycerol and silica\cite{Simon1926}. Soon after, similar results were reported in ethanol\cite{Kelley1929} and in the following years studies of excess entropy in glasses were extended to organic compounds\cite{Greet1967,Bestul1964,Tammann1930}, ionic melts, metallic alloys\cite{Chen1976} and so on. Furthermore,  Langer and  Sethna showed that the excess entropy usually obtained from the specific heat during heating (cooling) only provides an upper (lower) bound for the entropy\cite{Langer1988}. In spite of all previous results, in the late 1990s the debate on the reality of the excess entropy in the glass state was renewed in several works \cite{kivelson1999metastable,reiss2009apparent,gupta2009configurational}, based on the incapability to treat metastable states by thermodynamics and statistical thermodynamics. However, as addressed by Goldstein and Johari\cite{goldstein2008reality,goldstein2011reality,johari2011mechanical,johari2011specific}, and recently concluded by Schmelzer and Tropin\cite{schmelzer2018glass}, the nonexistence of the excess entropy in the glass state is in disagreement with the absolute majority of experimental and theoretical investigations of this process and the nature of the vitreous state.\\
\indent In order to explain this phenomenon Gibbs and DiMarzio\cite{Gibbs1958} proposed a theoretical description based on a lattice polymer model, in which below the so-called glass transition temperature ($T_g$) the liquid is frozen-in in a single configuration and unable to explore other configurations. This idea led to the potential energy landscape (PEL) description articulated originally by Goldstein\cite{Goldstein1969} as a topographic viewpoint of condensed phases and later formalized by Stilliger and Debenedetti\cite{Debenedetti2001}. In the framework of the PEL it is possible to separate the entropy, at low temperatures, in two contributions. One part which is configurational, arises from the exploration of different basins and the other, which is vibrational, originates from intrabasin thermal motions\cite{Stillinger1998,FrankH.Stillinger1999}. In the Gibbs-DiMarzio description there are not significant changes in the intrabasin vibration spectrum assuming that the excess entropy in glasses is entirely configurational. Nevertheless, in Goldstein's viewpoint the excess entropy has contributions from atomic and molecular vibrations. In this description, the excess entropy decreases linearly due to the linear dependence with temperature of the vibrational part, while the frozen-in configurational entropy remains constant below $T_g$.\\
\indent Although the phenomena of the glass transition relies on dynamics, a link between thermodynamics and dynamics is made through the Adam-Gibbs (AG) equation that relates the excess entropy to the relaxation time\cite{angell2002specific,kapko2008thermodynamics}. The driving force behind the structural relaxation would be the configurational entropy gained by the system as it explores distinct inherent structures\cite{berthier2019}. The frozen-in configurational entropy of the glass contains the information about the number of basins that are accessible to the supercooled liquid just prior to the glass transition.\\    
\indent In this work, we apply an efficient methodology based on non-equilibrium methods to separate the vibrational and configurational entropies in the binary Cu$_{50}$Zr$_{50}$ and ternary Cu$_{46}$Zr$_{46}$Al$_{8}$  bulk metallic glasses (BMGs) by means of molecular dynamics simulations. We choose the aforementioned metallic alloys because their properties are well known and reported in several experimental and theoretical studies\cite{altounian1982,buschow1981thermal,zhalko1994electronic,fan2006thermophysical,ZhouJCP2015,ZhangJAC2012}. Recently, the vibrational entropy contribution of these alloys was obtained experimentally using direct \textit{in situ} measurements of the vibrational spectra allowing separation of the vibrational and configurational contributions of entropy in BMGs\cite{Smith2017}. Here, we employ a purely thermodynamic methodology to compute the entropy. Our computational methodology is applied to calculate and split up the entropy of a BMG into configurational and vibrational contributions using a realistic interatomic potential.
\section{METHODS}
\subsection{Simulation setup}
%
\indent We used a simulation cell containing 4000 atoms. Periodic boundary conditions were employed to avoid surface effects. The interatomic interactions were modeled using an embedded atom method (EAM) potential as given in Ref. \cite{cheng2011atomic}. The simulations are performed using the molecular dynamics open code LAMMPS\cite{Plimpton1995}, with a time step of $\Delta t = 1fs$. The temperature and pressure are controlled using the Langevin thermostat and the Nos\'{e}-Hoover pressure barostat, with external pressure $P = 0$, and damping parameters $\tau_L =1fs $ and $\tau_{NH} =1ps $, respectively.\\
\subsection{Protocol and methods}
%
\indent Entropy, as well as free energy, are thermal variables, i.e., they depend on the entire accessible volume in the phase space. Thus, the calculations of these variables require special methods. In particular, for atomistic simulations, several methods are available to obtain these quantities\cite{Rickman2002}. In this work we used the adiabatic switching (AS)\cite{Watanabe1990} and reversible scaling (RS) \cite{DeKoning1999,DeKoning2001} methods to obtain the absolute free energy as a function of temperature. Both methods provide an accurate estimation of the free energy, including all anharmonic effects.\\
\indent Two reference systems were used during the AS simulations in order to obtain the absolute free energy, i.e., a collection of harmonic oscillators or the Einstein crystal (EC) and the Uhlenbeck-Ford model\cite{PaulaLeite2016,Leite2015} (UFM), for solid and liquid phases, respectively. In order to obtain the initial configurations, the system was equilibrated at $T = 1800$ K in the liquid phase during $1$~ns, right after it was quenched to $300$ K using a fixed cooling rate of $100$~K/ns and finally equilibrated again at this temperature during $1$~ns. $T_g$  was estimated in a similar manner to that done in Ref. \onlinecite{PhysRevMaterials.3.085601}, being $623$ and $713$~K for binary and ternary alloys, respectively. These results should be compared with the experimental findings\cite{ZhouJCP2015} of $664$~K for the binary alloy and $701$~K for the ternary alloy. These discrepancies between calculated and experimental results for $T_g$, typical of these type of calculations\cite{ZhangJAC2012}, are due to the very high cooling rate used in the simulations, finite-size effects, and limitations of the interatomic potentials. From the constant pressure specific heat of Cu$_{50}$Zr$_{50}$ that we obtained using this cooling rate, one can determine $T_g$ as the temperature at which the specific heat, after dropping from the peak, begins to decrease very slowly, exhibiting a behavior in good quantitative  agreement with the experimental specific heat of the glass\cite{Smith2017} for temperatures below $T_g$ (see Appendix A).\\
\indent At the first stage, we obtain the initial absolute free energy $G(T_0)$ of the alloy by means of the AS method. Here, $T_0$ stands for the temperature at which the reference system is used in order to obtain $G(T_0)$. In the solid phase, $T_0 = 300$ K and the reference system is the EC. It is important to note that $G(T_0)$ calculated using the EC includes only vibrational contributions for a given initial atomic configuration. In the liquid phase, on the other hand, $T_0 = 1800 $ K and we use as a reference system the UFM. This is an ultra-soft and purely repulsive pairwise interaction potential which resembles a liquid-like behavior\cite{PaulaLeite2016,Leite2019}. Thus, we calculated $G(T_0)$ using the AS formula: $G(T_0) = G_0 + W_{AS}$, where $G_0$ is the free energy of the reference system and $W_{AS}$ is the work done during the AS process. Since the work done during AS is calculated dynamically, a systematic error (SE) is generated during the process. Notwithstanding, if the switching process is performed slowly enough, within the linear response regime, the SE is eliminated changing $W_{AS}$ by the quasi-static work $\bar{W}$, obtained as\cite{DeKoning1997} $\bar{W} = (W^{dyn}_{for}-W^{dyn}_{back})/2$, where $W^{dyn}_{for}$ is the $W_{AS}$ done during the AS simulation from the alloy to the reference system and $W^{dyn}_{back}$ is the $W_{AS}$ performed in the inverse process.\\
\indent Once $G(T_0)$ is obtained, we use these values as references to calculate the free energy of the alloy in a wide range of temperatures using the RS method. The deduction of the RS equation can be found in several references \cite{DeKoning1999,DeKoning2001,Miranda2004}. Here, we only present the final result,
\begin{equation}
\text{G}(T) = \frac{\text{G}(T_0)}{\lambda} + \frac{3}{2}Nk_BT_0\frac{\ln\lambda}{\lambda} + \frac{W(\lambda)}{\lambda}, 
\label{RS1}
\end{equation}
where $\lambda$ is a scaling parameter defined as $\lambda = T_0/T$, and $W$ is the external work done when the scaling factor $\lambda$ is changed from $1$ to $T_0/T$. This work is estimated as $W = \int_{1}^{\lambda_{f}}\frac{d\lambda}{dt}\: U_\text{EAM}(\mathbf{\Gamma}(t))dt$, where $\mathbf{\Gamma}(t)$ is the vector in the phase space containing the information of all coordinates and momenta as a function of time. Thus, using Eq. $(\ref{RS1})$, the absolute free energy is obtained from $T_0$ to a final temperature $T (= T_0/\lambda_f$). Since the work done is calculated dynamically, a dissipation is generated during the process and the corrected quasi static work is obtained similarly to $\bar{W}$ calculated during the AS method.\\
\section{RESULTS AND DISCUSSION}
\subsection{Free energy}
%
\indent Because the calculation of the absolute free energy for solid systems using the EC as a reference is commonly called the Frenkel-Lad method\cite{frenkel1984new} we use $G_{FL}$ for the vibrational free energy obtained by means of RS in the solid state and, by analogy we use $G_{UF}$ for the free energy obtained from UFM as a reference for the liquid state.\\
\indent Fig.~1(a) depicts the absolute vibrational free energy of the glass ($G_{FL})$ and the absolute total free energy of the liquid  ($G_{UF})$, as functions of temperature, for the binary Cu$_{50}$Zr$_{50}$ alloy. Both curves in Fig.~1(a) were obtained from an average over 10 independent RS simulations. The value of G$_{UF}^0$ obtained for the liquid state at 1800 K agrees well with those previously reported in Refs.~\cite{Harvey2011,Leite2019} for the same percentage of Cu and Zr. Since we are dealing with glasses and glass transition, the cooling rate plays an important role in the application of the RS method. In order to guarantee a fixed cooling rate, the $\lambda$ parameter must vary as (see Appendix B) $\lambda = 300/(300 + \kappa t)$, where $\kappa$ is the desired cooling rate. Fig. 1(b) shows the behavior of $\lambda$ and the temperature (inset) as a function of time. Thus, using this functional form the system is always quenched at the same cooling rate. The free energy curves of the ternary alloy are included in Appendix C.\\ 
\indent For specific details of how to perform the AS-RS simulations in LAMMPS see Refs.~\cite{Freitas2016} and \cite{Leite2019} for the solid and liquid phases, respectively.\\
\indent In order to validate the methodological aspects of our calculations, such as the EAM interatomic potential, and the AS and RS methods, we estimated the melting point of the B2 crystalline phase of the Cu$_{50}$Zr$_{50}$ alloy (see Appendix D).
\begin{figure}[h!]
	\vspace*{-0.5cm}
	\centering
	\includegraphics[width=\linewidth,height=110mm]{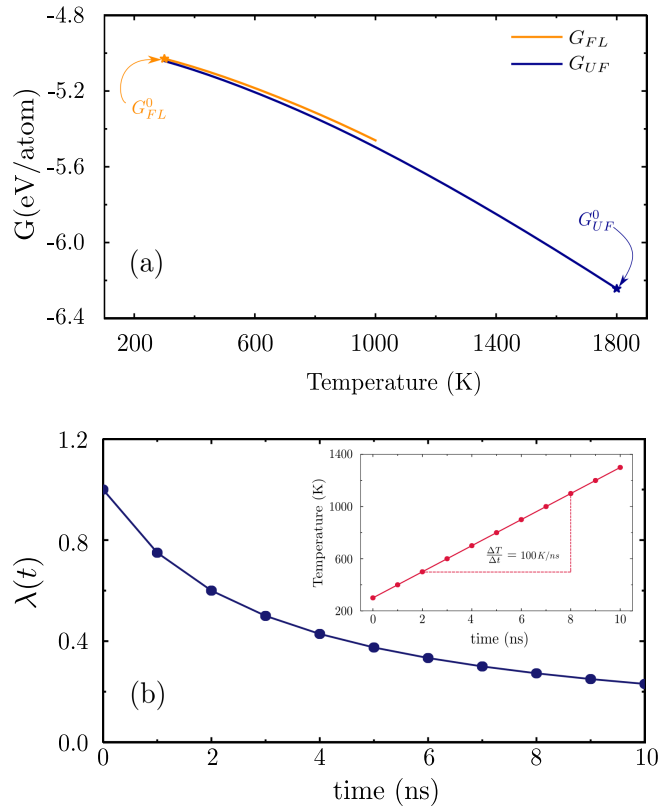}
	\caption{(a) Absolute Gibbs free energy per atom of the Cu$_{50}$Zr$_{50}$ metallic alloy. $G^0_{FL}$ and $G^0_{UF}$ are $G(T_0)$ obtained by means of AS using EC and UFM, respectively. (b) $\lambda$ parameter and temperature behavior as a function of time during RS simulation.}
	\label{figure1}
\end{figure}
\subsection{Splitting up entropy}
\indent The entropy is obtained by means of numerical calculation of $S=-(\partial G/\partial T)$. Since $G_{FL}$ contains only the vibrational contribution of the glass, $S_{FL}$ garnered through it only contains the vibrational part of the glass entropy. On the other hand, $S_{UF}$ is the total entropy of the liquid (or of the glass for temperatures below $T_g$) obtained from $G_{UF}$. $S_{UF}$ becomes the total entropy of the glass for $T<T_g$.\\ 
\indent Small statistical fluctuations in the free energy results are enhanced by the numerical differentiation, and therefore the Saviztky-Golay smoothing filter, performing a polynomial regression of third order, was used for noise reduction in the results for the entropy. In Fig. 2 we display the behavior of  S$_{UF}$ and S$_{FL}$ for the Cu$_{50}$Zr$_{50}$ and Cu$_{46}$Zr$_{46}$Al$_{8}$ alloys. A notable feature of the S$_{UF}$ curves for both alloys is a more rapid increase of their derivative below 1000~K, followed by a reduction of the derivative for temperatures approaching $T_g$, which results in a broad peak in the constant pressure specific heat c$_P$. This is a common feature of supercooled BMG forming liquids (see Appendix A). Below $T_g$, S$_{UF}$ and S$_{FL}$ decrease essentially at the same rate with temperature, thereby producing a frozen-in configurational contribution to the entropy. \\
\begin{figure}
	\centering
	\includegraphics[width=\linewidth,height=55mm]{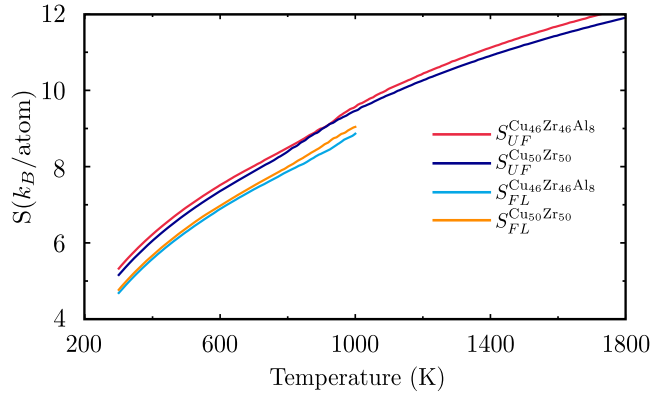}
	\caption{Total entropy $S_{UF}$ and vibrational part $S_{FL}$ of the binary Cu$_{50}$Zr$_{50}$ and ternary Cu$_{46}$Zr$_{46}$Al$_{8}$ alloys}
	\label{fig2}
	\vspace*{-0.4cm}
\end{figure}
\indent The configurational entropy is obtained from the difference $\Delta S_{LG}=S_{UF}-S_{FL}$ for temperatures below $T_g$. This behavior is depicted in Fig. 3; for $T<T_g$ the difference $\Delta S_{LG}=S_{conf}$ displays very small fluctuations around the average value, attaining an essentially constant value independent of the temperature. It is important to note that the configurational entropy of the Cu$_{46}$Zr$_{46}$Al$_{8}$ glass is 70\% higher than that of the Cu$_{50}$Zr$_{50}$ glass. This occurs because, as can be seen from Fig. 2, $S_{UF}^{\text{Cu}_{46}\text{Zr}_{46}\text{Al}_{8}}>S_{UF}^{\text{Cu}_{50}\text{Zr}_{50}}$, since the presence of an additional chemical element enhances both chemical and structural disorder, and also from Fig. 2, $S_{FL}^{\text{Cu}_{46}\text{Zr}_{46}\text{Al}_{8}}< S_{FL}^{\text{Cu}_{50}\text{Zr}_{50}}$, because the glass containing Al has a smaller atomic volume than that of the other glass (see Appendix E), leading to weaker anharmonic effects. The total configurational entropy $NS_{conf}$, where $N$ is the number of atoms, is related to the number of different structures that the glass can assume by $\Omega_{conf}=e^{NS_{conf}/k_B}$. Thus, within the framework of the PEL, the addition of aluminum to the alloy significantly increases the number of basins accessible to the supercooled liquid, which can be estimated to be $\Omega_{conf}^{\text{Cu}_{46}\text{Zr}_{46}\text{Al}_{8}} \sim (\Omega_{conf}^{\text{Cu}_{50}\text{Zr}_{50}})^{1.7}$, immediately prior to the glass transition. The factor 1.7 comes from the 70\% increase in the configurational entropy. Thus, the larger the number of basins, the greater the number of configurations the supercooled liquid can access and the more effective the relaxation of the liquid would be, and therefore, the easier it would be to form a more stable glass. By considering the Adam-Gibbs relation, one can see that this analysis is consistent with the experimental results by Zhou \textit{et al.}\cite{ZhouJCP2015} that show that just prior to the glass transition, the viscosity of liquid Cu$_{46}$Zr$_{46}$Al$_{8}$ is lower than that of liquid Cu$_{50}$Zr$_{50}$. It is well known that Cu$_{46}$Zr$_{46}$Al$_{8}$ has a higher glass-forming ability (GFA) than that of Cu$_{50}$Zr$_{50}$.\cite{ZhouMaterialsTransactions2010} A glass with higher configurational entropy is a less-ordered system than one with lower entropy. The paper by Wang \textit{et al.}\cite{WangJAP2008} displays results of x-ray diffraction experiments for both alloys, which show that glassy Cu$_{50}$Zr$_{50}$ exhibits diffraction peaks that are related to crystalline phases of the alloy, whereas in glassy Cu$_{46}$Zr$_{46}$Al$_{8}$ these peaks are absent. The lack of crystalline order inhibits crystallization and is related to the higher GFA of Cu$_{46}$Zr$_{46}$Al$_{8}$. Our results suggest that although other factors may be at play, configurational entropy can be helpful to understand the effect of the addition of a minor-alloying element on the GFA of BMGs.\\
\begin{figure}
	\centering
	\includegraphics[width=\linewidth,height=55mm]{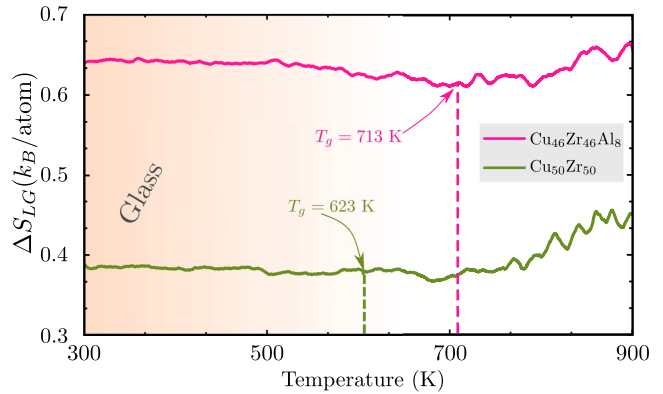}
	\caption{Entropy difference $S_{UF}-S_{FL}$ for the binary Cu$_{50}$Zr$_{50}$ and ternary Cu$_{46}$Zr$_{46}$Al$_{8}$ alloys}
	\label{figure2}
	\vspace*{-0.2cm}
\end{figure}
\subsection{Comparison with experiment}
\indent In a recent work, Smith \textit{et al.}\cite{Smith2017} experimentally separated the configurational and vibrational entropies of the same alloys using \textit{in situ} neutron diffraction and differential scanning calorimetry. Their findings were obtained by determining the vibrational entropy of the glass and the crystal of the alloys in a range of temperatures of about 100 K below $T_g$. Smith \textit{et al.} found the configurational entropy of the Cu$_{50}$Zr$_{50}$ glass to be 0.27~$k_{B}/atom$, while we have determined 0.39~$k_{B}/atom$; we think there is a quite reasonable agreement between our results and the experimental findings, in particular if one takes into account that the configurational entropy in this case is a small quantity, resulting from the difference between two numerically similar quantities, namely, S$_{UF}$ and S$_{FL}$. We determined for the Cu$_{50}$Zr$_{50}$ glass a vibrational entropy of 7.0~$k_{B}/atom$ at 600~K, which is in fair agreement with the experimental value of 6.3~$k_{B}/atom$ at that temperature. We have found the vibrational entropy of the Cu$_{46}$Zr$_{46}$Al$_{8}$ glass to be 6.9~$k_{B}/atom$ at 600~K. However, in Ref.~\cite{Smith2017}, the vibrational entropy results for the two glasses are not directly compared due to technical difficulties, and because of that we do not compare our results for the glass containing Al with the respective experimental findings.\\
\indent Smith and co-workers found that the vibrational entropy of the glasses is almost equal to that of their crystalline counterparts for that interval of temperatures, concluding that the excess entropy, i.e., the difference of entropy between crystal and liquid phases for temperatures below $T_g$, is entirely configurational. In order to compare our results with Smith's experimental work, we calculate the vibrational entropy of the crystalline phase of the Cu$_{50}$Zr$_{50}$. However, Cu-Zr metallic alloys have a complex crystalline structure, which has been described by Kalay \textit{et al.}.\cite{kalay2010devitrification,kalay2011high} Upon heating, the glass undergoes devitrification into crystallites of three coexisting crystalline phases: orthorombic Cu$_{10}$Zr$_{7}$, tetragonal  CuZr$_2$, and cubic CuZr(B2). According to Kalay,\cite{kalay2010devitrification}  these crystallites have dimensions just under 1$\mu m$, resulting in a very complex structure at the nanoscale. This crystalline structure is so complex that is impossible to simulate it. Nevertheless, in Fig. 4, we compare the behavior of the vibrational entropy of each crystalline phase with the vibrational entropy of the glass of the Cu$_{50}$Zr$_{50}$ alloy. The entropies of the crystals differ from the vibrational entropy of the glass by a small amount (0.2--0.4~$k_{B}/atom$ at 600~K), as compared with the magnitude of the entropies themselves (6.6--7.0~$k_{B}/atom$ at 600 K), in contrast to the findings by Smith \textit{et al.} A possible explanation for this discrepancy is that in their case, the crystal is a mixture of crystalline phases and there could be significant anharmonic contributions to the vibrational entropy coming from the interfaces between the crystallites, which would increase the vibrational entropy. It should be emphasized that these interfaces exist at the nanoscale, therefore, they can give rise to substantial anharmonic effects. This explanation is corroborated by the work of Ohsaka \textit{et al.}, \cite{OhsakaAPL1997} who found that the difference between the thermal-expansion coefficients for the glass and the crystal (not a single crystal) of the pentanary alloy Zr$_{41.2}$Ti$_{13.8}$Cu$_{12.5}$Ni$_{10.0}$Be$_{22.5}$ is small, which means that the anharmonic effects in both glass and crystal should be similar.\\
\indent We have also studied the excess entropy of the glass with respect to the crystalline phase CuZr$_2$. Our results, which are in agreement with the description proposed by Goldstein,\cite{Goldstein1969} are presented in Appendix F.\\
\begin{figure}
    \vspace*{-1.cm}
	\centering
	\includegraphics[width=\linewidth,height=55mm]{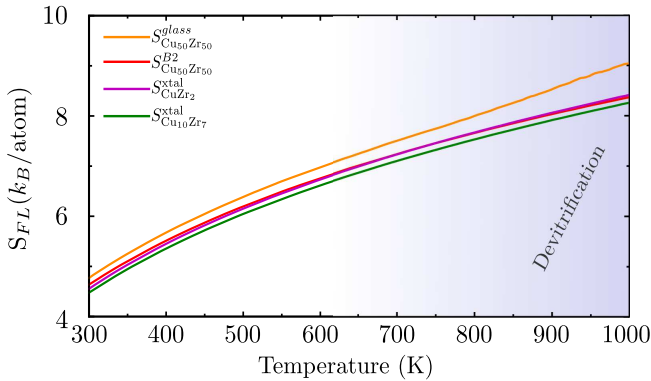}
	\caption{ Vibrational entropy of Cu$_{50}$Zr$_{50}$ (glass), Cu$_{10}$Zr$_{7}$, CuZr$_2$, and CuZr(B2) as a function of temperature. For temperatures above of $T_g$, the glass turns into a liquid and the entropy S$_{FL}$ is no longer only vibrational, but also contains a configurational part.}
	\label{figure3}
\end{figure}
\section{SUMMARY}
%
\indent We have separated the vibrational and configurational contributions to the entropy of two metallic glasses, namely, Cu$_{50}$Zr$_{50}$ and Cu$_{46}$Zr$_{46}$Al$_{8}$, through molecular dynamics simulations. We employed in our calculations a robust methodology, whose qualities are evidenced by the good agreement between the results of our simulations and the experimental available data \cite{Smith2017} for the vibrational and configurational entropies of the Cu$_{50}$Zr$_{50}$ metallic glass. The main advantage of this methodology is that one can separate the two contributions to the entropy without comparing the results for the glass with those for the crystal.
We determined the configurational entropy of the glass Cu$_{46}$Zr$_{46}$Al$_{8}$ to be about 70\% higher than that of the other glass Cu$_{50}$Zr$_{50}$. Configurational entropy is directly related to the number of distinct configurations that the glass can assume and, within the PEL framework, to the number of basins available for the supercooled liquid. Thus, just prior to the glass transition, the larger the number of basins, the more effective the relaxation toward a more stable glass would be. Since it is well known that Cu$_{46}$Zr$_{46}$Al$_{8}$ has a larger GFA than that of Cu$_{50}$Zr$_{50}$, our findings suggest that, although other factors may be at play, glass configurational entropy can be useful to study the effect of the addition of a minor-alloying element on the GFA of BMGs.\\
\section*{ACKNOWLEDGMENTS}
\indent We gratefully acknowledge support from the Brazilian agencies CNPq, CAPES, under Project No. PROEX-0487, and FAPESP under Grants No. 2010/16970-0 and No. 2016/23891-6. The calculations were performed at CCJDR-IFGW-UNICAMP. We thank J. Rego and B. Peluzo for the kind explanation of the crystalline structure and the valuable discussions. Finally, we thank R. Paula-Leite for the helpful discussion about UFM and the free energy calculation in the liquid phase.\\ 
\appendix

\section{HEAT CAPACITY}
\indent Fig. 5 shows the specific heat of the Cu$_{50}$Zr$_{50}$ alloy obtained using two procedures, namely, by calculating the numerical derivative of the enthalpy and by computing the enthalpy fluctuations. Except for the height of the peak in the specific heat, the results yielded by both procedures agree very well. As explained before, $T_g$ is considered to be the temperature at which the specific heat, after dropping rapidly from the peak, starts to decrease rather slowly, assuming values in good quantitative agreement with the experiment\cite{Smith2017} for temperatures below $T_g$. The value of $T_g$ that is obtained is essentially the same as that previously determined by the temperature at which the kink in the enthalpy curve occurs\cite{PhysRevMaterials.3.085601}.
\begin{figure}[h!]
	\centering
	\includegraphics[width=\linewidth,height=55mm]{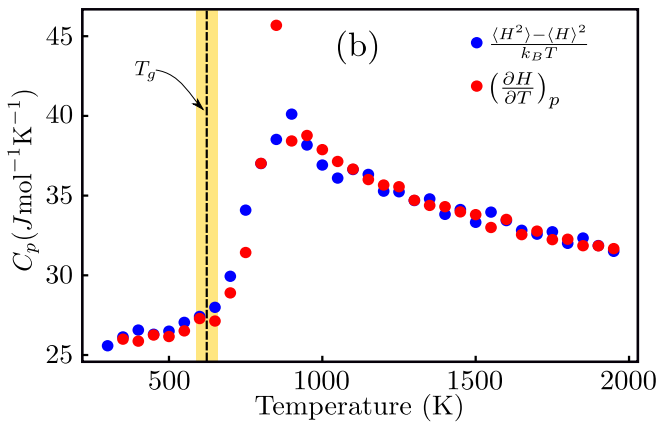}
	\caption{Constant pressure heat capacity $C_p$ of Cu$_{50}$Zr$_{50}$ using two methods: derivative of the enthalpy and enthalpy fluctuations obtained from our molecular dynamics simulations.}
\end{figure}
\section{RS-SIMULATIONS AT A CONSTANT COOLING RATE}
In computational studies of the glass transition, the cooling rate in which the system was quenched is a fundamental quantity, since the glass transition depends on it. Therefore, when the RS method is used, in order to keep the cooling rate fixed we consider the differential equation
\begin{equation}
\frac{dT}{dt} = \kappa,
\end{equation}
where $\kappa$ is the constant cooling rate desired. Since RS is used to obtain the free energy in a wide range of temperatures given by $T =T_0/\lambda$, we need to solve Eq $(1)$ with the conditions $T_0 = 300 K$ and $\kappa = 100 K/ns$. Thus, we have
\begin{equation}
\int_{1}^{\lambda} d\left(300/\lambda^{'}\right) = \int_{0}^{t} 100 \:\: dt^{'}.
\end{equation}
The solution of this equation provides the functional form for $\lambda(t)$ as:
\begin{equation}
\lambda(t) = \frac{300}{300 + 100 t}.
\end{equation}
Using Eq.~(B3) we guarantee the cooling rate to be fixed at $100K/ns$. We are interested in obtaining the free energy as a function of temperature in the interval $[300K-1300K]$ and, therefore,
\begin{equation}
1300 K = \frac{300}{\lambda},
\end{equation}
Then we have to vary $\lambda$ from $1$ to $0.23$ in order to obtain the free energy in the desired interval. The entire simulation time required is obtained as:
\begin{equation}
t_{sim} = \frac{1300K-300K}{100K/ns} = 10ns,
\end{equation}
since we are using a time step of $1fs$, and so we need $10^7$ molecular dynamics time steps to reach $\lambda(t_{sim}) = 0.23$. Figure 1(b) depicts the behavior of $\lambda$ and $T$ during the switching process.
\section{FREE ENERGY OF THE TERNARY ALLOY}
Fig.~6 depicts the free energy results for the Cu$_{46}$Zr$_{46}$Zr$_{8}$ alloy. The free energy of the liquid is given by the $G_{UF}$ curve and the $G_{FL}$ curve gives the vibrational contribution to the glass.
\begin{figure}[h!]
	\centering
	\includegraphics[width=\linewidth,height=55mm]{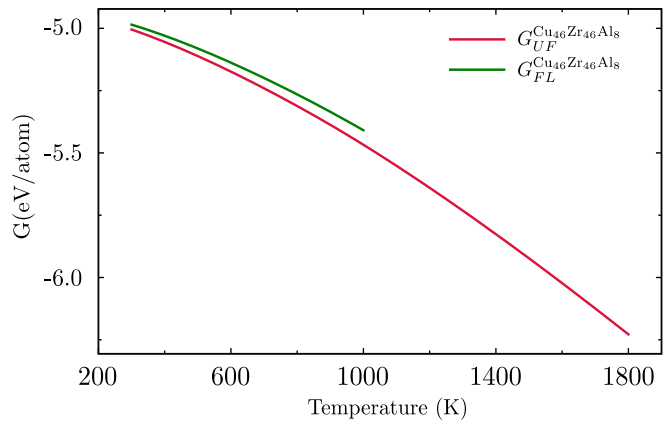}
	\caption{Free energy per atom of the Cu$_{46}$Zr$_{46}$Zr$_{8}$ metallic alloy. The $G_{UF}$ and $G_{FL}$ curves depict the free energy of the liquid and the vibrational free energy of the glass, respectively.}
\end{figure}
\\%
\section{MELTING POINT}
It is well known that the Cu$_{50}$Zr$_{50}$ alloy crystallizes into a B2 structure when it is cooled slowly enough. In Fig. 7, we present the results of a calculation of the melting point obtained using the RS method. The crossing between the free energy curves for the crystal and the liquid phases gives the thermodynamic melting point, which was found to be approximately 1316 K; this value differs from the experimental melting point (1210 K) by about 100 K. This difference is due to the fact that the melting temperature is extremely sensitive to small changes in the free energy. All other calculations using empirical potential, analytical, and \textit{ab initio} procedures provide the melting point with a relative error with respect to the experiment of approximately 8$\%$ \cite{carvalho1980,Gunawardana2014}.\\
\begin{figure}[h!]
	\centering
	\includegraphics[width=\linewidth,height=55mm]{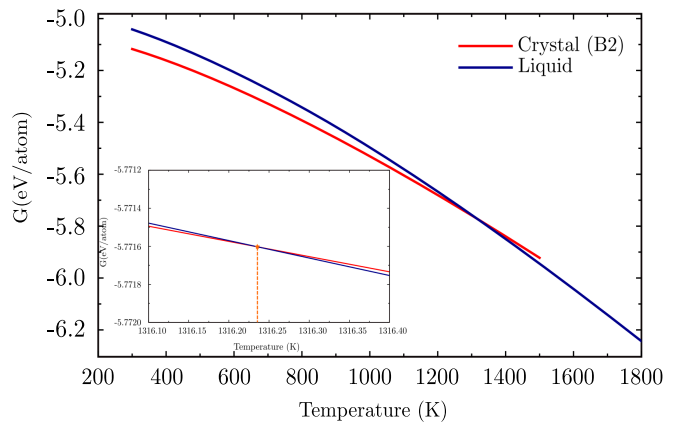}
	\caption{Free energy per atom of the crystal (B2) and liquid phases of Cu$_{50}$Zr$_{50}$ metallic alloy. The inset shows the crossing point between crystal and liquid free energy curves.}
\end{figure}
\section{MOLAR VOLUME}
In Fig. 8, we show the molar volume of both alloys as a function of temperature. It is interesting to note that at very high temperature (liquid phase), both alloys have very similar molar volumes, however, at low temperature (glass phase), the molar volume of the Cu$_{46}$Zr$_{46}$Zr$_{8}$ alloy is remarkably lower than that of the Cu$_{50}$Zr$_{50}$ alloy, considering that the amount of aluminum introduced in the alloy is quite small.
\begin{figure}[h!]
	\vspace*{0.4cm}
	\centering
	\includegraphics[width=\linewidth,height=55mm]{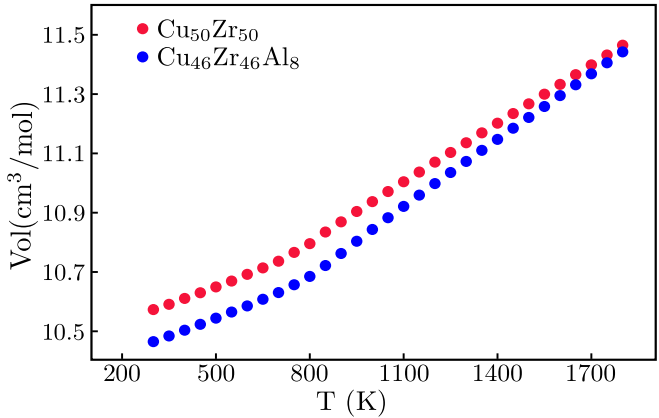}
	\caption{Molar volume of Cu$_{50}$Zr$_{50}$ and Cu$_{46}$Zr$_{46}$Zr$_{8}$ metallic alloys.}
\end{figure}\\
\section{EXCESS ENTROPY}
\begin{figure}
	\centering
	\includegraphics[width=\linewidth,height=55mm]{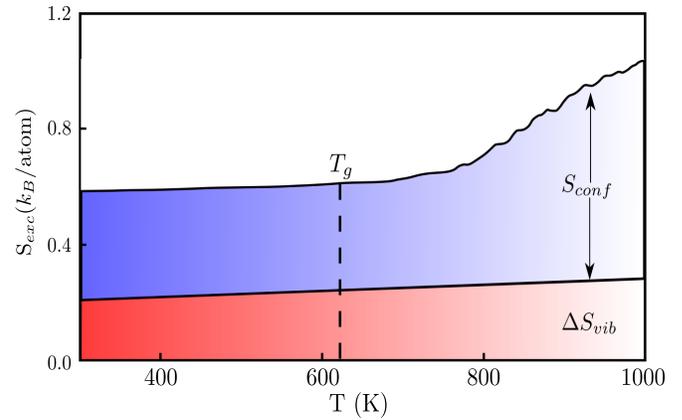}
	\caption{ The excess entropy as a function of temperature. Since the vibrational contributions of glass and CuZr$_2$ are not the same, the excess entropy decays linearly when the temperature decreases suggesting a behavior as described by Goldstein.}
	\label{figure4}
\end{figure}
We have estimated the excess entropy of the Cu$_{50}$Zr$_{50}$ alloy with respect to three different crystalline phases, namely, Cu$_{10}$Zr$_{7}$, CuZr$_2$, and CuZr. Fig. 9 depicts the S$_{exc}$ with respect to one of them, CuZr$_2$. The excess entropies with respect to the other two crystalline phases exhibit similar behavior.\\
Our results for the excess entropy of the alloy with respect to single-crystal structures show that, because of the slight difference between the vibrational entropies of the glass and the crystal, the excess entropy increases linearly with temperature for temperatures below $T_g$, which is in accord with Goldstein's description of the excess entropy \cite{Goldstein1969}, instead of that of Gibbs and de DiMarzio \cite{Gibbs1958} observed in the experiment \cite{Smith2017}. This discrepancy is possibly due to anharmonic effects in the crystal used as reference in the experiment, which is a mixture of crystalline structures. In Fig. 9, the curve depicting the excess vibrational entropy for temperatures above $T_g$ is an extrapolation of the curve for temperatures below $T_g$.

\bibliographystyle{apsrev4-2}
%
\end{document}